\documentclass[journal,hideappendix]{vgtc}        
\usepackage{graphicx}

\onlineid{0}

\vgtccategory{Research}

\vgtcpapertype{please specify}

\title{Investigating How Music Affects Persuasion, Engagement, and Emotion in Data Videos}

\author{%
  Sarmistha Sarna Gomasta, Mahmood Jasim, Hossein Hadisi, Yvonne Jansen, Pierre Dragicevic,\\
  Narges Mahyar, Ali Sarvghad
}

\authorfooter{
  \item
  Sarmistha Sarna Gomasta is with the University of Massachusetts Amherst.
  E-mail: sgomasta@umass.edu.

  \item
  Mahmood Jasim is with Louisiana State University.
  E-mail: mjasim@lsu.edu.
  
  \item
  Hossein Hadisi is an independent researcher.
  E-mail: hosseinhadisi@yahoo.com.

    \item
  Yvonne Jansen is with Inria, CNRS, University of Bordeaux, and LaBRI.
  E-mail: yvonne.jansen@cnrs.fr.

  \item
  Pierre Dragicevic is with Inria, CNRS, University of Bordeaux.
  E-mail: pierre.dragice@gmail.com.

  \item
  Narges Mahyar is with City St George’s, University of London.
  E-mail: narges.mahyar@city.ac.uk.

  \item
  Ali Sarvghad is with City St George’s, University of London.
  E-mail: ali.sarvghad@citystgeorges.ac.uk.
}

\abstract{%
Data videos have become a prominent vessel for communicating data to broad audiences, and a common object of study in information visualization. Many of these videos include music, yet the impact of music on how people experience data videos remains largely unexplored. We conducted a preregistered study into the effect of music across three dimensions: persuasion, engagement, and emotion. We showed online participants an existing data video (1) without any music, (2) with its generic default music, and (3) with custom music designed by a professional composer. We found that the default music helped make the data video more persuasive. However, the effects of custom music were more mixed, and we did not find that music increased engagement. In addition, and contrary to our expectations, our participants reported more intense emotions without music. Our study contributes new insights into the intersection of music and data visualization and is a first step toward guiding designers in creating impactful data-driven narratives.

}


\graphicspath{{figs/}{figures/}{pictures/}{images/}{./}} 

\usepackage{tabu}                      
\usepackage{booktabs}                  
\usepackage{lipsum}                    
\usepackage{mwe}                       
\usepackage{csquotes} 
\usepackage{dirtytalk}
\usepackage{mathptmx}                  
\usepackage{balance}
\usepackage{cite}

\usepackage{comment}

\begin{document}



\newcommand{\nm}[1]{{{\color{teal}{[\textbf{nm (#1)}]}}}}

\firstsection{Introduction}
\maketitle

Data videos are motion graphics that combine visual and auditory stimuli to present and communicate data dynamically \cite{amini2015understanding}. 
Through the use of data-driven graphics such as information visualization~\cite{khan2011data} and infographics~\cite{naparin2017infographics}, they serve as a method for engaging and instructing a wide range of audiences by rendering data in a manner that is 
engaging, cognitively accessible, and informative.
For instance, during the global pandemic, the World Health Organization (WHO) and similar institutions leveraged data videos to raise awareness and communicate information about the pandemic including prevention and recovery strategies~\cite{GovernmentOfCanada2021PhysicalDistancing}.
Similarly, data videos have been shown to be an effective vehicle for education as well as raising awareness around global and social issues~\cite{sakamoto2022persuasive, amini2015understanding}.

To persuade viewers through engaging content that sustains user attention, and ensures content clarity, data video designers have explored a plethora of techniques for creating data videos ranging from diverse narrative constructs~\cite{cao2020examining} and audio-visual components~\cite{shen2023data} to cinematic aspects such as framing, camera effects, and cinematic beginnings~\cite{xu2022wow}. 
Additionally, a substantial amount of such data videos are accompanied by some form of background music~\cite{cao2020examining}. 
Music has long been a universal language that transcends cultural and linguistic barriers to evoke emotions, inspire change, and deliver critical messages ~\cite{merriam1964anthropology}. 
Across different types of multimedia content, music has been shown to have a profound impact on the viewers' subjective and psychological interpretation of a scene~\cite{galan2009music}. 
Outside of entertainment, economics research has shed light on the impact of background music on consumers' behavioral changes~\cite{galan2009music}, such as influencing purchase decisions~\cite{areni1993influence}, increasing time spent shopping~\cite{milliman1982using} or enhancing customer loyalty~\cite{morris1998effects}. In addition, music has been explored as a tool for influencing memorability and learning performance~\cite{martin2015effectiveness}, as well as affecting decision-making in various scenarios, such as risk-taking in financial markets or compliance with public policy~\cite{north2008social}.

Despite the well-studied impact of music in multimedia content, little attention has been paid to investigating the effect of music accompanying data videos. 
As of yet, the effect of music in data videos and its functions and impact on the viewers remain underexplored. 
In this work, we explore this knowledge gap by investigating three key dimensions related to the impact of music on data videos -- persuasion, engagement, and emotion. 
We extend prior work that showed the importance of persuasion~\cite{sakamoto2022persuasive}, viewer engagement~\cite{amini2018hooked, xu2022wow}, and emotion ~\cite{campbell2019feeling} in data visualization to investigate the impact of music in these critical dimensions. To this end, we address the following research questions:
\begin{itemize}[noitemsep,leftmargin=*]
    \item \textbf{RQ1}: How does music in data videos affect the \textit{persuasive power} of data videos? 
    \item \textbf{RQ2}: How does music in data videos affect viewers' \textit{engagement}?
    \item \textbf{RQ3}: How does music in data videos affect viewers' \textit{emotions}?
\end{itemize}

Considering the interdisciplinary nature of these research questions, we assembled a team with expertise in visualization, music, social computing, and social activism. We then began by creating a collection of publicly available data videos, where we included data videos from online journalism, popular video platforms, blogs, and previous visualization research~\cite{amini2015understanding, amini2016authoring, amini2018hooked}. 
Our inclusion criteria for selecting the data videos included music, data visualization and/or infographics, elements that promote empathy-building or provoke emotions, and effective message communication (as judged by the team of authors). 
We excluded videos with audio narration to isolate the impact of music without the influence of spoken content. 
We explored a variety of global, contemporary, stimulating, and thought-provoking topics. The topics we considered include but are not limited to climate change, gun violence, poverty, income inequality, gender inequality, education inequality, water crises, food waste, and cyberbullying. 
Six members of our research group viewed, analyzed, and ranked the videos to select the video we used for this study after three rounds of deliberations. 
The selected video was on the global water crisis, which contained rich visualizations, had clear data and facts, was relatable for a broad audience, and had the potential to evoke emotions~\cite{water2024}. 
With the help of a professional music composer in our team, we created three variations of the data video for our study -- (1) \textit{no music}, (2) with the \textit{default music} accompanying the data video, 3) with \textit{custom music} intended to amplify some of the messages included in the data video, thereby replacing the default music.

We conducted a between-subject study on Amazon Mechanical Turk (257 participants) to compare the three variations of the data video. 
We found evidence that the data video with default music leads to more persuasion,  although the evidence is much less clear for custom music.
We found no evidence that default music is more engaging than no music and even found evidence that custom music is less engaging than no music.
In addition, not only do we lack evidence that music evokes more emotions, but we also have some evidence to the contrary.
Our study is the first of its kind to examine how music influences viewers of data videos.
It contributes to the field by delving into the intricate relationship between music and visualization in data videos. 

\section{Background}

This section discusses prior works investigating persuasion, engagement, and emotion in visualization studies. 
We reviewed the design and relevant tools in the data video from our research direction.
Additionally, we highlight prior studies focusing on music, its impact as well as music in visualization or any related fields.

\subsection{Data Videos}
In the last few decades, conveying data and disseminating knowledge through visual and audio narration has seen major successes in communicating scientific discovery~\cite{siregar2022discovery}, persuading informed decision-making~\cite{wang2016animated}, enabling sensemaking~\cite{yang2022towards}, and uncovering complex dynamics~\cite{islam2019overview}. 
There exists a growing body of research on crafting data videos and how they may help people understand complex issues or persuade them to consider different viewpoints~\cite{segel2010narrative}. 
For instance, Amini et al. analyzed 50 data videos and identified the common narrative structures (i.e., establisher, initial, peak, and release), visualization types, and attention cues, critical for data communication~\cite{amini2015understanding}. 
Others have explored guidelines for creating data videos with a prominent focus on animation techniques. 
Examples include detailing common animation types~\cite{amini2018hooked}, animated transitions focusing on smooth context switching and fluent narratives~\cite{tang2020narrative}, animated visual narrative strategies~\cite{wang2016animated}, and identification of design primitives of animated data graphics~\cite{thompson2020understanding}.
Previous works also provide higher-level design guidelines on augmenting data videos without incurring conflict with the original content~\cite{tang2020design}, establishing common narrative patterns~\cite{yang2021design} and cinematic styles~\cite{xu2022wow}, as well as taxonomies of narrative and visuals~\cite{cao2020examining}.  
These guidelines have allowed data videos to be established as a vehicle for communicating information to a broader audience for education as well as raising awareness of local, social, and global issues in an engaging, accessible, and informative manner. 

\subsection{Persuasion in Data Visualization}
Prior works showed how visualizations can influence people's data interpretation and decision-making~\cite{pandey2014persuasive,garreton2023attitudinal}.
For example, cartographic visualization is often used in politics and persuasion by influencing certain viewpoints~\cite{kent2016political}. 
Across various domains, designers often use more gratuitous graphs or charts to deliver information that they want to highlight~\cite{cairo2019charts}.
Others have explored how information can be visualized effectively to communicate negative information while promoting positive outcomes through data stories~\cite{lan2022negative}. 
Prior research also suggests that graphical aspects of visualizations, such as aspect ratio~\cite{beattie2002impact}, visual encodings~\cite{correll2017black}, color choices~\cite{bartram2017affective}, highlighting text or annotations~\cite{franciscani2014annotation}, title~\cite{kong2018frames}, visual embellishments~\cite{bateman2010useful}, framing and slants~\cite{hullman2011visualization} can encourage certain takeaway messages by persuading the viewers. 

Beyond visualization in general, many researchers explored the design of persuasive data videos. 
For instance, Choe et. al designed persuasive data videos by augmenting four persuasive elements --- primary task, dialogue, system credibility, and social support for self-tracking feedback~\cite{choe2019persuasive}. 
Sakamoto et al. proposed that including affect-centered narratives in data video can lead to positive outcomes and influence viewers' attitudes towards contact-tracing apps~\cite{sakamoto2022persuasive}.
Sallam et al. emphasized the interaction between the health-related data video's content and the viewer's affective states~\cite{sallam2022towards}. 
They also found that the viewer's personality is interlinked with persuasion, showing that highly neurotic individuals are less likely to change their opinion than any other personality.
In effect, data videos have been shown to be a powerful tool in influencing people's attitudes and behaviors in societal contexts (e.g., climate change \footnote{\url{https://www.youtube.com/watch?v=ffjIyms1BX4}}, health ~\cite{choe2019persuasive}). 
While prior research shows various methods and techniques to increase the persuasive power of data videos, the impact of music on data videos remains yet underexplored. 

\subsection{Engagement in Data Visualization}
Engagement plays a critical role in data visualization~\cite{mahyar2015towards}, as it is often perceived as a measurement for the effectiveness of visualizations~\cite{burns2020evaluate}. 
Prior research has explored elements that impact viewers' engagement with visualizations. 
For instance, Borkin et. al, have prioritized visual and semantic associations alongside visual artifacts~\cite{borkin2013makes} to achieve viewers' attention.
Researchers also explored narrative modes to make the viewers' experiences meaningful by achieving narrative intents~\cite{ojo2018patterns}. 
Others have explored ways to increase engagement using colors~\cite{bartram2017affective}, mood~\cite{lan2021smile}, animation transitions~\cite{heer2007animated}, visual embellishments~\cite{bateman2010useful}, affect~\cite{saket2016beyond}, sensory experiences~\cite{kwok2019gaze} and social or contextual factors~\cite{kennedy2016engaging}.  
Engagement is often measured in factors such as memorability, times spent, enjoyment, and user interactions~\cite{mahyar2015towards}. 
Mahyar et al. presented a five-degree scale --- expose, involve, analyze, synthesize, and decide --- to measure user engagement~\cite{mahyar2015towards}.

Amini et. al have refined the engagement scale for data videos by covering five factors --- affective involvement, enjoyment, aesthetics, focused attention, and cognitive involvement --- to measure viewers' engagement~\cite{amini2018hooked}.
In data videos, various attributes may play vital roles in designing an engaging and interactive experience for viewers, such as visual aesthetics, attention, and control. 
Furthermore, animations in data videos are a popular choice for information delivery~\cite{amini2018hooked}, heightening focused attention, and keeping viewers engaged in the video. 
In addition, prior works also suggest that photographic representation provides lucid insights, evokes feelings, and makes data videos appealing to viewers~\cite{o2010development}. 
For instance, Xu et. al., presented six cinematic styles and twenty-eight guidelines to create an attractive beginning of data video to engage viewers~\cite{xu2022wow}. 
They extended their work to design four types of punctuation mark endings through the lens of cinematic styles for better viewer engagement~\cite{xu2023end}. 
As an alternative, Herath et al. generated nine design implications for robotic presentation in data video to engage the audience through gestures and expressions~\cite{herath2023exploring}. 
In summary, engagement in visualization is multifaceted, including elements such as visual aesthetics, narrative modes, affective involvement, and sensory experience, which require further inquiries to identify the interplay among these elements. 

\subsection{Emotion in Data Visualization}
Despite the undervaluation of emotions in data visualization literature~\cite{lan2023affective}, an increasing number of studies have explored emotions in visualization in an effort to understand how they impact viewers~\cite{lan2021smile,wang2019emotional}. 
Emotion is observed as a common factor in artistic visualization~\cite{wang2019emotional}, data stories~\cite{shi2022breaking}, influencing decision-making~\cite{garreton2023attitudinal} or behavioral responses~\cite{lee2014m}, and achieving cognitive intents~\cite{bateman2010useful}. 


Researchers explored affective data visualization to disseminate data to a larger audience~\cite{lan2023affective}. 
For instance, the U.S. Gun Death is one of the popular projects that sparked the necessity to include emotions in visualization for social good~\cite{boy2017showing}.
To incorporate emotions in visualizations, design techniques such as anthropographics are widely used to humanize specific individuals~\cite{morais2020showing}, make data more realistic~\cite{rost2017data}, and focus on certain data points~\cite{morais2020showing}. 
The design elements of affective techniques have been shown to attract viewers' attention, evoke emotions, and provide more engaging experiences~\cite{campbell2019feeling}. 
Researchers have also focused on visual elements such as anthropomorphic characters~\cite{boy2017showing}, proximity techniques~\cite{campbell2019feeling}, framing of data~\cite{hullman2011visualization}, visual styles~\cite{bateman2010useful}, or label visualization with affective nouns~\cite{lan2021smile}. 
Others have demonstrated how music, when synchronized with visualizations, can evoke emotions by providing an auditory stimulus of music and visual content~\cite{ronnberg2019musical}. 

\subsection{Music and its effects on the viewers}
Music is extensively studied as a complementary factor to the visual modality in supporting comprehension of visual representations~\cite{krygier1994sound}. Due to its pervasive nature, music has been demonstrated to influence human behavior in different domains such as prosocial activities~\cite{kniffin2017sound}, different decision-making tasks (e.g., perceptual, lexical, value-based)~\cite{perez2022background}. For instance, strategic music in war-related documentaries might lead people to make humanitarian aid decisions or fair donations~\cite{shao2023composing}. 
A selective music might reduce the fear of needles and promote blood donation campaigns (e.g., Keep the World Beating)~\cite{tan2022song}. 
Apart from social change, studies have demonstrated that background music, tempo, and genre can significantly influence customers' purchasing decisions~\cite{areni1993influence}, affecting what consumers purchase and total time spent in a shop~\cite{milliman1982using}. 
For example, slower-tempo music has been shown to encourage shoppers to move more leisurely, leading to increased sales~\cite{milliman1982using}. 
Additionally, the choice of music can reinforce brand identity and enhance customer loyalty~\cite{morris1998effects}. 
Music is also a powerful vessel for evoking emotions and creating connections, often becoming a civil rights anthem to strengthen social movements~\cite{tan2019social}. 
Additionally, music can influence narratives, such as highly markable dramatic sounds typically used in fiction films to form a sense of tension, create empathy with certain characters, or emphasize particular narratives~\cite{rogers2014music}. For instance, in Jaws, John Williams music defined the character of sharks as a menace and fearful \footnote{https://www.soundstripe.com/blogs/the-psychology-of-music-for-film}.

Despite music's impact on influencing viewers, little focus has been given to how music affects viewers of data videos. 
While prior studies have focused on the design guidelines, and multiple structural components to measure data video's efficacy~\cite{cao2020examining, amini2018hooked}, to the best of our knowledge, no research measured the effects of music in data videos. 
The major challenges for measuring the effectiveness of music in data videos stem from the ubiquity of highly tailored music in our everyday lives increasing peoples' sonic awareness~\cite{quinones2016ubiquitous} and varied preferences of music and alignments on the visualization topic~\cite{yang2013mozart}. 


In this work, we investigate three key dimensions related to the impact of music on data videos --- persuasion, engagement, and emotion. 

\section{Selecting Data Videos}
In this section, we describe the process of the data video topic and video selection process for our study. 
We also describe the pilot study and lessons learned that influenced our final video selection. 

\subsection{Topic Selection}
Our goal was to select a contemporary and relevant topic that would be relatable and comprehensible for a broad audience~\cite{johnson2005communication}. To achieve this, we explored a range of topics including global or social crises, politics, sports, science, education, mythology, magnitude comparison, social media, finance, and arts. We examined data videos that were mentioned or investigated in prior works~\cite{amini2015understanding, cao2020examining}. In addition, we explored contemporary news articles, current affairs, and the zeitgeist to identify thought-provoking topics. We have excluded any highly polarized topics (e.g., anti-vaccination, gun control) because people usually overlook evidence that contradicts their mindset~\cite{pandey2014persuasive}, especially in highly polarized socio-moral issues. We have also focused on topics that are easily comprehensible and can engage an audience without overwhelming them~\cite{pandey2014persuasive}.



\subsection{Data Video Dataset}
\label{DC}
Our data curation process for the study involved two stages. Initially, we curated 40 data videos from popular video platforms such as YouTube, Vimeo, online blogs, journalism, and previous data video research~\cite{amini2015understanding, amini2016authoring, cao2020examining}. 
To find data videos across these sources, we used various keywords, such as ``data stories'', ``short documentary'', ``animated statistics'', ``data video'', and ``infographics'' to search for related videos.
To focus on the effect of music and avoid potential confounding effects~\cite{djikic2011effect}, we excluded data videos where the music had lyrics or spoken or voice-over audio narrations which may in themselves already create physiological responses~\cite{richardson2020engagement} or influence persuasive purposes~\cite{zoghaib2019persuasion}. 
We also excluded non-English data videos.

At the end of the filtering process in two stages, our final data video collection list had 73 data videos. 
The complete list of our data video collection is included in the OSF.

\subsection{Preliminary Video Selection}
\label{Preliminary-Video-Selection}
From our data video dataset, we primarily identified 40 data videos as candidates for the study based on factual information visualization~\cite{amini2015understanding}, visual aesthetics~\cite{xu2022wow,xu2023end}, narrative frameworks~\cite{cao2020examining}, and the length or duration of the video~\cite{guo2014video}. 
Prior works also suggest short videos (up to 3 minutes) can influence the users to hold their attention better~\cite{guo2014video}. 
We also considered videos with higher view counts, as they are more often likely to provoke audience interest~\cite{huang2020good}. 
Three members of our research team independently scored the data video on a likelihood scale from 1 (lowest score) to 5 (highest score) based on the abovementioned criteria. 
Based on their scores, we selected five different data videos on a diverse set of topics namely climate change, overpopulation, alcohol's effect, gender equality, and human population evaluation. 
Our goal was to explore videos across different topics to test our hypotheses across various topics to identify generalizable results. 

\subsection{Pilot Study}
We used the five selected data videos to run a pilot study. The role of the pilot study was to first test the effect of music on a range of measurements including persuasion, comprehension~\cite{burns2020evaluate}, engagement, and emotion (see \autoref{Ms} for details). We also included open-ended questions about participants' experiences of watching the videos with or without music. The study was fully between-subject, meaning that each participant saw one of the five data videos, either without its default music (no music condition) or with its default music (music condition). Thus, we had a total of $5 \times 2 = 10$ experimental conditions. The study was hosted on Qualtrics\footnote{https://www.qualtrics.com//}. We used snowball sampling \footnote{https://methods.sagepub.com/foundations/snowball-sampling} to recruit participants using word of mouth on social media and emails. A total of 79 participants were randomly assigned to one of the 10 experimental conditions (38 saw a video without music and 41 saw a video with music). Participants took an average of 30 minutes to complete the study.
Results were largely inconclusive regarding the effect of music on any of the dependent variables measured. 
However, we found evidence that participants reacted differently to different videos. For example, they demonstrated a positive attitude change on average after having watched the gender inequality video, but a negative attitude change after having watched the human population video, irrespective of the music condition. 

The inconclusive results obtained in the pilot study were likely due to a lack of statistical power arising in part from the variability between videos. We saw that the choice of video had a large impact on how people experience the video (e.g., in terms of attitude change, emotion, or comprehension), and this impact was likely much larger than the impact of music alone. This is consistent with previous studies suggesting that a range of video characteristics such as duration~\cite{guo2014video}, framing~\cite{hullman2011visualization}, and choice of music~\cite{ansani2020soundtracks} have an impact on how people experience videos. As a consequence, including different videos in a study adds variability that makes it more difficult to detect the effect of music alone. Looking at the effect of music on each of the individual videos is also difficult without a very large sample size (in our pilot, only about 16 participants were assigned to a particular video). Therefore, one important change we made to our study design was to include a single data video. The use of a single video makes the study less generalizable, but it reduces its complexity and maximizes statistical power.
Furthermore, the majority of the participants mentioned that the study involved too many questions and it was difficult for them to concentrate. To address survey fatigue~\cite{jeong2023exhaustive}, we removed comprehension from our measurements, which seemed to be the most noisy measure in our pilot.

\subsection{Final Data Video Selection}
\label{Final video selection}
To select one data video, we updated our video selection criteria (\autoref{Preliminary-Video-Selection}) based on the lessons learned from our pilot study. Our refined criteria were that the data videos should be rich in facts, contain visualization~\cite{bateman2010useful}, and include affective intent (e.g., call for action~\cite{lee2022affective}) to bring attention to an issue or persuade viewers. 
This led to our second set of data video collection processes, where an additional 33 data videos were collected from YouTube, Vimeo, online blogs, journalism, and previous data videos research~\cite{amini2015understanding, amini2016authoring, amini2018hooked}. 
Six members of our research team individually ranked the videos with an appropriateness scale from 1 (highest rank), and 5 (lowest rank). 
Based on the ranking, we finalized a set of six data videos on topics namely cyberbullying, the water crisis (two videos), US gun control, injuries, and global health issues. 
We performed a last round of ranking on the same appropriateness scale from 1 (highest rank), and 5 (lowest rank) to finalize one data video from these six candidate data videos. 
Based on this process, we picked a data video on the global water crisis.
The original video was on YouTube and developed by Animaker~\cite{water2024}\footnote{\url{https://www.youtube.com/watch?v=HSENz13ZFxo}}. 
The video was 1 minute and 22 seconds long and included factual information about the global water crisis presented using multiple visualizations and animations, and a call to action to reduce the waste of water. 

\section{Study Design}
\label{sec:study-design}
We conducted a between-subject study with three conditions: no music, default music, and custom music. The following section provides a detailed description of our study. The study design and its details including measurements, sample size, participant selection criteria, data exclusion criteria, and our hypotheses were preregistered on OSF before data collection\footnote{Link to preregistration: \url{https://osf.io/wk854/}}. Our study materials including stimuli, questionnaires, data, and analyses are also available on OSF\footnote{Link to the OSF project: \url{https://osf.io/adxhv/}}.

\subsection{Measurements}
\label{sec:measurements}
For all our primary and secondary measurements, we used 25 questions on a 7-point scale, which we internally coded from -3 to +3. The full questionnaire is available on our OSF project.

\label{Ms}
\paragraph{\textbf{Persuasion.}}
Following Pandey et al.~\cite{pandey2014persuasive} approach, we captured persuasion by measuring \textbf{attitudinal persuasion}, which is the difference in attitude on a given topic before and after having watched the video. Attitudinal persuasion was determined by measuring:

\begin{itemize}[noitemsep,leftmargin=*]
    \item \textbf{Pre-video attitude:} We asked three questions measuring participants' attitudes toward the water crisis discussed in the video before they watched the video. The values were re-coded so that higher values indicate higher agreement with the video's message. All the questions were worded in the same way: ``\textit{To what extent do you agree that}...'', followed by topic-specific statements. For instance, \textit{``To what extent do you agree that limiting plastic usage can help address the water crisis?}''
    
    \item \textbf{Post-video attitude:} Participants answered the same three questions measuring their attitude after having watched the video.
\end{itemize}

We also measured persuasion through three secondary variables: behavioral persuasion, believability, and general persuasive potential.

\begin{itemize}[noitemsep,leftmargin=*]
    \item \textbf{Behavioral Persuasion:} As persuasion is highly related to behavioral intentions~\cite{boy2017showing}, we measured behavioral persuasion as a secondary variable with one question stating ``\textit{To what extent would you likely take action to address the water crisis.}''.
    
    \item \textbf{Believability:} As persuasion and believability are related to each other~\cite{thomas2019can}, we asked a question to measure the participant's believability of the information presented in the video. Our question is one single 7-point scale question stating ``\textit{To what extent do you think that the information in the video is reliable}?''.

    \item \textbf{General Persuasive Potential:} In the user-centered design process, the general persuasive potential works as a subjective tool for measuring a system's capability to persuade individuals ~\cite{choe2019persuasive}. To measure whether our data video is persuasive or not, we also asked three relevant questions with the following format: ``\textit{To what extent do you...}'' followed by a topic-specific statement.
\end{itemize}

\paragraph{\textbf{Engagement.}}

We assessed participant's engagement through 4 questions based on prior studies: participant's interest in video and topic~\cite{hung2017assessing}, interest to learn more about the topic~\cite{wittenberg2021minimal} and sharing intention~\cite{ahmed2021learning}. 
We also included a question about topic familiarity to assess if it was a moderator of engagement.

\paragraph{\textbf{Emotion.}}
Batson's list of emotion adjectives~\cite{batson1987prosocial} is widely used to measure emotion, empathetic concern, and personal distress in visualization~\cite{boy2017showing}. We selected a subset of eight emotions that are relevant to our video's topic and to keep our study short: sympathetic, moved, compassionate, alarmed, grieved, upset, worried, and disturbed. We omitted emotions such as tender, warm, and softhearted as these emotions don't correspond with our selected topic. The participants reported how strongly they felt each emotion on a 7-point scale ranging from ``extremely'' (+3) to ``not at all'' (-3). All the adjectives appeared in a randomized order to reduce the risk of potential order effects.

\paragraph{\textbf{Message emphasis.}}
\label{sec:message-emphasis}
We preregistered a secondary research question as ``\textit{Can emphasis added through music to one of the video's messages make that message stand out from the others}?''. To answer this question, we showed participants a list of five different messages that were conveyed in the data video at different moments (order randomized) (\autoref{tab:multinominal_messages}), and asked them which of the messages they found was best communicated. One of the messages (M2) was ``1 in 10 people have no access to water'', and the professional composer has put a strong emphasis on it in the custom music group. 

\paragraph{\textbf{Enjoyment of the Music.}}
\label{sec:opinion-music}
For both default music and custom music groups, we asked ``\textit{To what extent do you think the music in the data video is enjoyable?}'' on a 7-point scale ranging from `extremely enjoyable (+3)'' to ``extremely unenjoyable (-3)''. For the no music group, a different question has been asked on the same scale stating as -``\textit{To what extent do you think adding music to this data video would be effective?}'' ranging from ``extremely effective (+3)'' to ``extremely ineffective (-3)''.

\begin{table}[htbp]
\renewcommand{\arraystretch}{.8}
\caption{Selected key messages in the data video}
\label{tab:multinominal_messages}
\begin{tabular}{l p{0.75\columnwidth}}
\toprule
\textbf{Shorthand} & \multicolumn{1}{c}{\textbf{Message}} \\
\midrule
M1 & Only 1\% of all water is available for humans. \\
M2 & 1 in 10 people have no access to water. \\
M3 & Waterborne diseases kill a child every 15 seconds. \\
M4 & Washing dishes, flushing the toilet, or taking baths can waste lots of water. \\
M5 & 1 plastic bottle takes 8.4 liters of water to make. \\
\bottomrule
\end{tabular}
\end{table}

\subsection{Hypotheses}
In relation to the research questions we stated in our introduction, we formulated the following three hypotheses:

\begin{itemize}[noitemsep,leftmargin=*]
    \item \textbf{H1 -- Persuasion}: The data video with music (default and custom) will be more persuasive than the data video without music, as measured by our attitudinal persuasion scale.

    \item \textbf{H2 -- Engagement}: The data video with music (default and custom) will lead to higher engagement than the data video without music, as measured by our engagement scale.

     \item \textbf{H3 -- Emotion}: The data video with music (default and custom) will lead to more intense reported emotions than the data video without music, as measured by our emotion scale.
\end{itemize}

\subsection{Stimuli and Conditions}
Our study includes a single independent variable---music, with three conditions:

\begin{itemize}[noitemsep,leftmargin=*]
\item \textbf{No Music}: The participant watched a data video without any music.
\item \textbf{Default Music}: The participant watched a data video with its existing music.
\item \textbf{Custom Music}: The participant watched a data video with custom music designed by a professional composer.
\end{itemize}

Apart from the music, the video was identical across all three conditions. Below is the description of the data video and its three versions in more detail. 


\paragraph{\textbf{Video stimulus.}}

As we explained in~\autoref{Final video selection}, we chose to use a single data video in this study. The video we picked was about the global water crisis. While the video matched the criteria we initially agreed upon, the timing of the video turned out to be too fast to enable the viewer to comfortably read the different visualizations and the text included in the video. We therefore adjusted its timing by slowing the video down to 75\% of its original speed. This adjustment was chosen as an optimal balance between comfortable reading speed and natural-looking animations. To avoid changing the tempo or the pitch of the original music, we re-looped it, that is, we copied the repeating music motive to fill the additional length of the slowed-down video.
We also added a fade-out at the end of the modified music, as was also done in the original music. 
Our professional composer also confirmed that the modifications did not change the spirit of the original design.

When integrating the music with the edited video, a minor error was identified at the 0:22 mark. The video stated ``1 in 10 have no access to water'' accompanied by a pie chart labeled ``water 1'' and ``water 10.'' These labels should have read ``1'' and ``9'', and the smaller pie slice should have been labeled ``No water'' instead. We also found another minor ambiguity at the 1-minute mark, where the video stated that dishwashing uses 100 liters of water. To provide more contextual clarity and avoid the potential perception of cherry-picking or dishonesty, we modified the text to specify that the stated water usage for dishwashing (100 liters) refers to the scenario where the water is continuously running, which may not be a common practice for everyone. After the video editing, the video length had become 1 minute 42 seconds for all conditions. The three video versions are available in our OSF project \href{https://osf.io/adxhv/}{(\textcolor{blue}can be accessed here)}.

\paragraph{\textbf{Custom music design.}}


\begin{figure}
    \centering
    \includegraphics[width=\columnwidth]{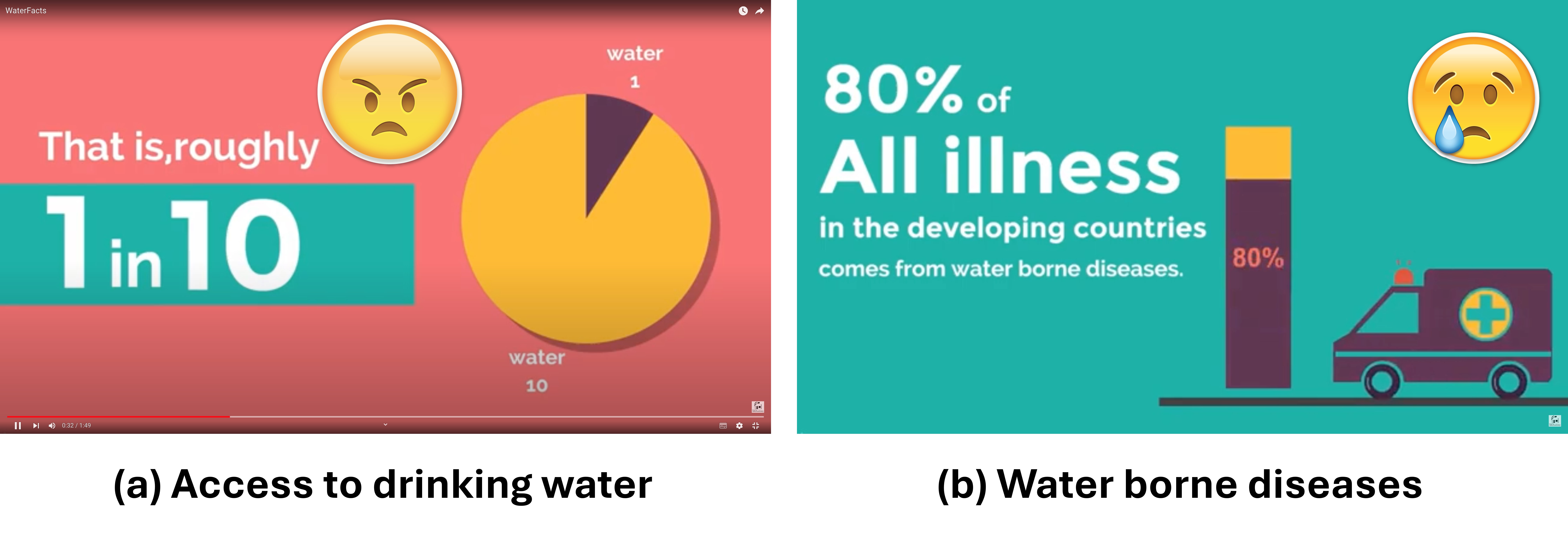}
    \caption{Emotion categorization in different messages}
    \label{fig:water_issues}
\end{figure}




The goal of our study was to explore the potential effect of music in influencing participants’ persuasion, engagement, and emotion. Sound is often used to complement data visualization and foster engagement with a captivating and compelling experience~\cite{krygier1994sound}. Prior research in related contexts such as sonification highlights that sound well-aligned with the data may convey meaningful information and evoke people's emotions~\cite{tsuchiya2015data}. 
After the pilot study, we made the strategic decision to introduce a third condition into our study, aiming to gain deeper insights into how different versions of music can affect viewers. 
By manipulating the type of music accompanying the data videos, we sought to unravel the nuanced effects that music can exert on the viewer's experience. However, manipulating the type of music needs a specific set of music knowledge, skills, and methods.

To design our third variation (custom music), the professional music composer on our team adopted a monologue-based approach, combining factual elements and emotional buildup to tailor the music based on the content. The composer envisioned a self-contained symphonic/sonata-like structure to meet certain emotional milestones similar to film music composing. To do this, he first identified important moments (i.e., conveying key facts or changes of tone), then annotated these moments using emojis and keywords to assign emotional categories to the respective frame of the video content (\autoref{fig:water_issues}). 
Then he used a standard composition tool (Logic Pro X) and applied well-known, occasional cliché ideas to present darkness, warning tone, excitement builds up, and suspension. 
For reaching peaks he used tools such as dissonance, extreme bass range, rhythmic development and subdivisions, and a combination of crescendo and accelerando (easing dynamics and tempo) respectively. More details are available on OSF.



\subsection{Sample Size and Participants}
\label{sec:pop}
Based on power analysis, we chose a sample size that would give us an 80\% probability of detecting a standardized effect size of $f$ = 0.2 (conventionally considered as a medium effect) \href{https://www.psychologie.hhu.de/fileadmin/redaktion/Fakultaeten/Mathematisch-Naturwissenschaftliche_Fakultaet/Psychologie/AAP/gpower/GPowerManual.pdf} {in an omnibus $F$-test} involving the three experimental groups (details in the preregistration). The required sample size was 82 participants per group, for a total sample size of 246 participants. 

We conducted our study on Amazon Mechanical Turk (MT). To ensure data quality~\cite{lu2022improving}, we recruited participants who have equal or more than 98\% Human Intelligence Task (HIT) approval rates with a number of HITs approved greater or equal to 500, and they needed to have Amazon Masters Certification in performing the tasks. On average, participants took 11 minutes to complete the survey and each participant was paid 10 USD through Amazon Payments.

\subsection{Procedure}

Our main study interface was hosted on Qualtrics. Participants accessed the interface by clicking a link in Amazon MT. They were first presented with an overview of the experiment, payment details, and contact information in a consent form.  They then began the study after agreeing to the informed consent form for participating in the experiment.

\begin{itemize}[noitemsep,leftmargin=*]
\item \textbf{Screening and demographics:} We screened potential participants using a pre-screening questionnaire including age, primary language, and auditory, and vision impairments. 
Participants in the ``default music'' and ``custom music'' groups were asked to turn their audio on and went through a basic audio check, where they listened to an audio recording of a number and typed the number they heard. If the answer was incorrect, the participant was asked to try again. If the answer was incorrect again, the participant was considered not eligible and the survey ended.

\item \textbf{Pre-video stage:} Participants in all three groups were given a brief introduction to the topic covered in the video: ``\textit{Many countries are experiencing a water crisis. We will ask you some questions about it, but you do not need to have prior knowledge about the topic. Be assured, there are no right or wrong answers to these questions}''. They answered the three pre-video attitude questions mentioned in \autoref{sec:measurements}, after which they were asked to watch the video that will follow carefully.

\item \textbf{Post-video stage:} After having watched the video, participants completed a post-video questionnaire consisting of twenty-one questions (see \autoref{sec:measurements}). They also answered an attention-check question, two questions related to the music track, and an optional open-text question to give feedback on the study.
\end{itemize}

Participants were not allowed to navigate between pages using the browser's forward/back buttons without answering questions, and could not modify their responses to the previous questions. Participants were only permitted to watch the video once but were allowed to pause, resume, and rewind the video by 10 seconds.

\subsection{Data Exclusion Criteria} 
We did not exclude any data based on the quality of the responses given.
However, we removed data from 14 participants who failed the audio check question after two attempts. We also removed data from 11 participants who failed the attention-check question. In addition, we excluded incomplete data from 21 participants, which included responses from any participant who used a browser or device other than the designated Qualtrics platform to watch the video or failed the captcha verification. 
Finally, we removed all duplicate responses. In total, 309 participants started the study, and 257 participant's responses (83\%) were considered valid and kept for further analysis.

\section{Analysis and Results}
\label{AR}

We collected valid data from 257 participants, for a planned sample size of 246 participants (see \autoref{sec:pop}). We had 85 participants in the no-music group, 85 in the default music group, and 87 in the custom music group. In this section, we report the results of our analyses, starting with our planned analyses.

\subsection{Planned Analyses}

The analyses reported here cover our three primary outcomes of interest: persuasion, engagement, and emotion intensity (see \autoref{sec:measurements} for details). They have all been preregistered (link in footnote of \autoref{sec:study-design}). The only changes to our preregistration are error fixes and a minor change in the way results are presented, the details of which can be found in OSF.

\subsubsection{General Methods}
With a sample size of $N$ = 85 to 87 subjects per condition and reasonably continuous measurement scales (37 possible values for attitude change, 28 for engagement, and 56 for emotion), it is reasonable to invoke the central limit theorem, stipulating that sampling distributions of sample means are approximately normal~\cite{baguley2018serious}. This assumption justifies the use of $t$-tests and $t$-distribution confidence intervals (CIs).

We report all effects using differences in means, 95\% CIs, and $p$-values. We, however, use an estimation approach to statistical inference, which emphasizes the examination of CIs reported graphically and sees statistical evidence as lying on a continuum rather than being binary~\cite{dragicevic2016fair}. We interpret $p$-values as continuous measures of evidence, using \textit{p}=.05 as a rough landmark instead of a strict cut-off~\cite{besanccon2019continued}. 

\subsubsection{RQ1 / H1 -- Persuasion}
\label{sec:attitude-change-results}

\begin{figure}
    \centering
    \includegraphics[width=\linewidth]{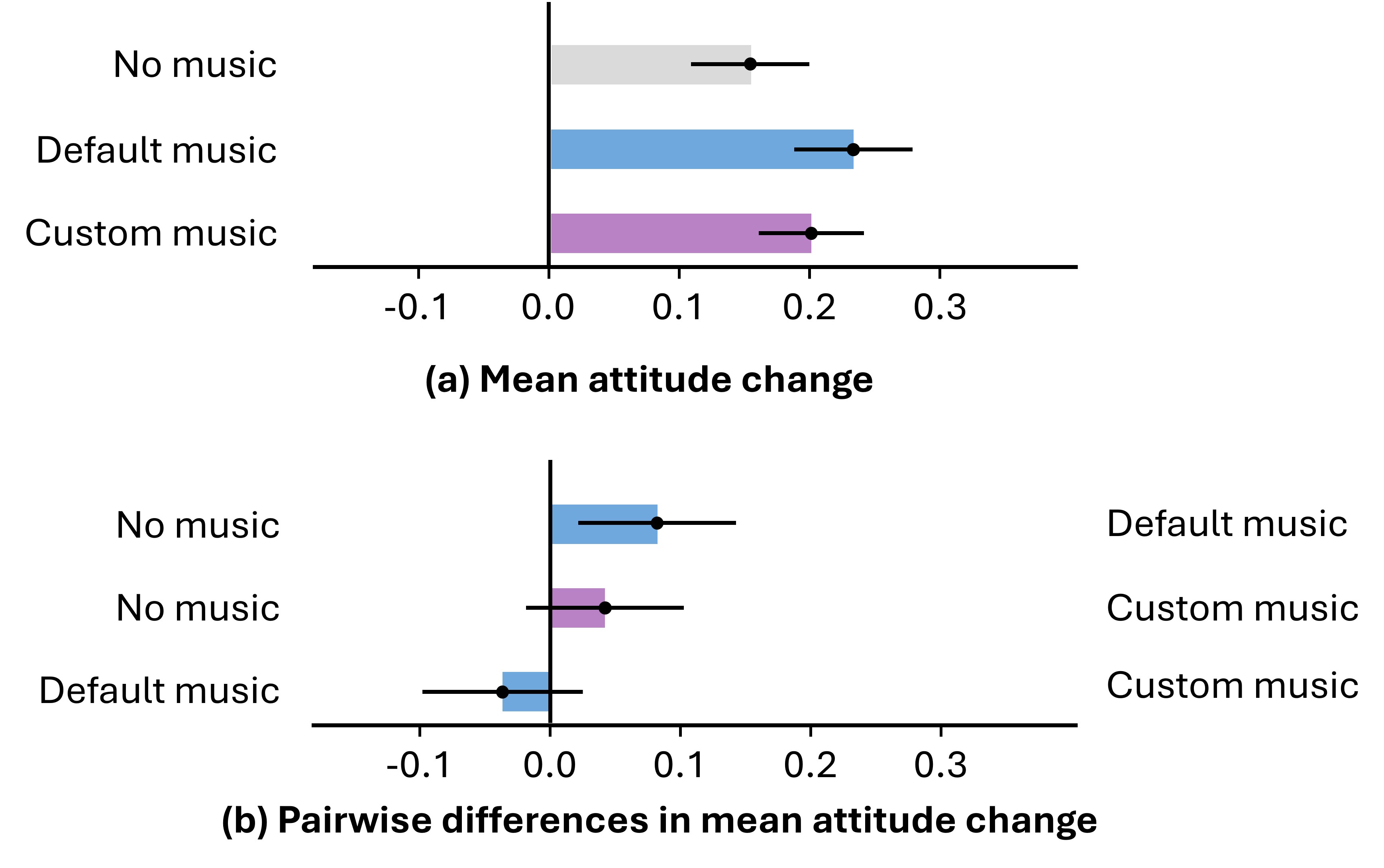}
    \caption{(a) Mean attitude change for each music condition. Attitude change is normalized such that a score of 0 indicates no change and a score of -1 or 1 indicates the maximum change possible. (b) Pairwise differences in mean attitude change between conditions (right condition minus left condition). Error bars are 95\% confidence intervals.}
    \label{fig:attitude_changes}
\end{figure}

\begin{table}
\renewcommand{\arraystretch}{0.8}
\caption{CIs and \textit{p}-values for differences in mean attitude change}
\label{table_example_persuasion}
\centering
\begin{tabular}{@{}lll@{}}
\toprule
Difference & 95\% CI & \textit{p}-value \\ \midrule
default music - no music & [0.017, 0.14] & .013 \\
custom music - no music & [-0.015, 0.11] & .14 \\
custom music - default music & [-0.095, 0.026] & .26 \\
\bottomrule
\end{tabular}
\end{table}

Our first research question was whether music can make data videos more persuasive. We hypothesized that our data videos with music (default and custom) would lead to more persuasion.

We computed the mean attitude change for each of the three experimental groups. Then, we computed the difference in mean attitude change between group pairs (i.e., default music vs no music; custom music vs no music; custom music vs default music). For each mean difference, we ran an independent-samples $t$-test, from which we derived a 95\% CI and a $p$-value. 

\autoref{fig:attitude_changes}-(a) shows the mean attitude change for each of the three conditions, with their 95\% CIs (plot not preregistered but included here for clarity purposes). It confirms that all three video versions lead to a clear positive attitude change on average. The default music condition appears to yield more attitude change, which is confirmed by the pairwise comparisons reported in \autoref{fig:attitude_changes}-(b) and \autoref{table_example_persuasion}. There is good evidence that the default music condition leads to more attitude change than the no music condition. The remaining results are less conclusive. It is possible that custom music also leads to more persuasion than no music, but the evidence is much weaker. Meanwhile, we did not find enough evidence for a difference between custom music and default music -- they likely differ, but we do not have enough data to determine the direction of the effect.

Conclusion: \textbf{The evidence overall supports H1}. We have good evidence that the data video with music leads to more persuasion, even though the evidence is much less clear for custom music.

\subsubsection{RQ2 / H2 -- Engagement}
\label{sec:engagement}

\begin{figure}
    \centering
    \includegraphics[width=\linewidth]{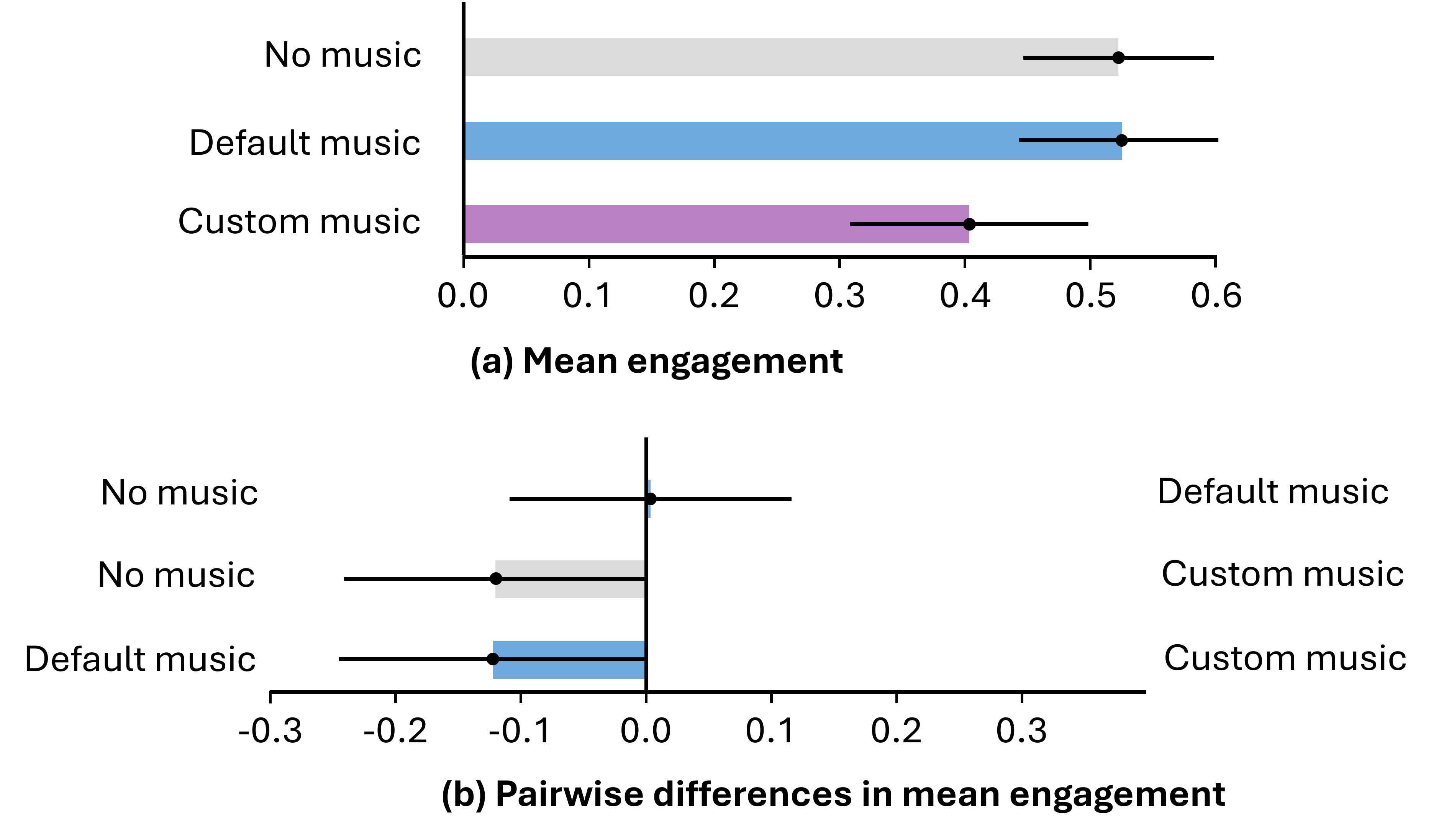}
    \caption{(a) Mean engagement score for each music condition. Engagement scores are normalized between -1 and 1, with a score of 0 indicating a neutral response to the engagement questions. (b) Pairwise differences in mean engagement between conditions (right condition minus left condition). Error bars are 95\% confidence intervals.}
    \label{fig:engagement}
\end{figure}

\begin{table}
\renewcommand{\arraystretch}{0.8}
\caption{CIs and \textit{p}-values for differences in mean engagement}
\label{table_example_engagement}
\centering
\begin{tabular}{@{}lll@{}}
\toprule
Difference & 95\% CI & \textit{p}-value \\ \midrule
default music - no music & [-0.11, 0.11] & .97\\
custom music - no music & [-0.24, 0.0020] & .054 \\
custom music - default music & [-0.24, 0.0023] & .055 \\
\bottomrule
\end{tabular}
\end{table}

Our second research question was whether music can lead to more engagement with data videos. We hypothesized that our data videos with music (default and custom) would lead to a higher engagement score. 

As before, we computed the mean engagement score for each of the three experimental groups and then computed the difference in mean attitude change between group pairs. We again computed 95\% CIs and $p$-values from independent-samples $t$-tests. 
As before, \autoref{fig:engagement}-(a) shows the mean engagement for each of the three conditions, confirming engagement is overall positive across all conditions (recall the full range is -1 to 1). Looking at the pairwise comparisons reported in \autoref{fig:engagement}-(b) and \autoref{table_example_engagement}, we have evidence that default music and no music lead to more engagement than custom music. 


Conclusion: \textbf{The evidence does not support H2}. We found no evidence that default music is more engaging than no music and even found evidence that custom music is less engaging than no music.




\subsubsection{RQ3 / H3 -- Emotion Evocation}

\begin{figure}
    \centering
    \includegraphics[width=\linewidth]{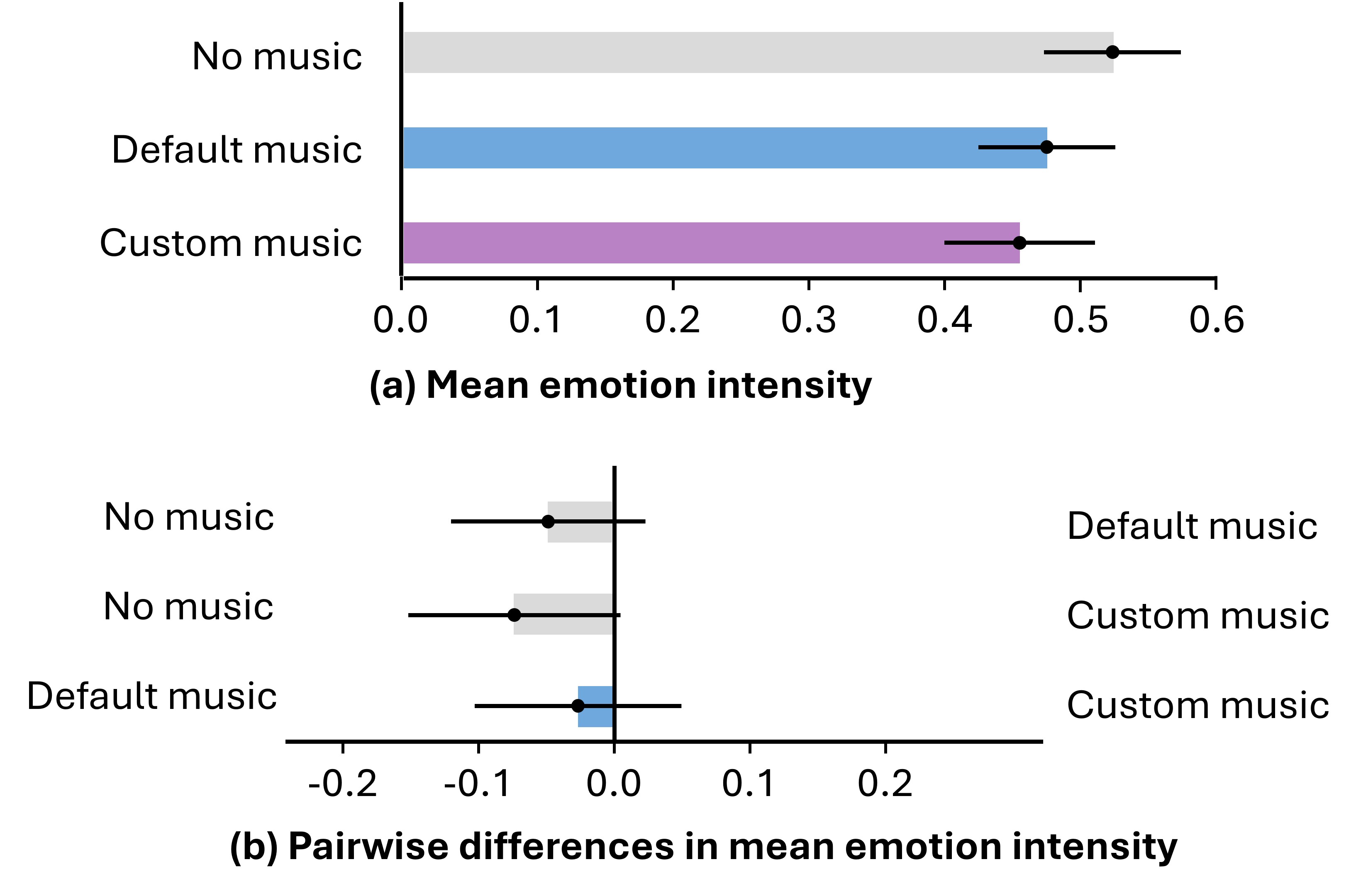}
    \caption{(a) Mean emotion intensity for each music condition. Emotion intensities are normalized between 0 and 1, with a score of 0 indicating no emotion, a score of 0.5 indicating moderate intensity, and a score of 1 indicating extreme intensity. (b) Pairwise differences (right condition minus left condition). Error bars are 95\% confidence intervals.}
    \label{fig:emotion}
\end{figure}

\begin{table}
\renewcommand{\arraystretch}{0.8}
\caption{CIs and \textit{p}-values for differences in mean emotion intensity}
\label{table_example_emotion}
\centering
\begin{tabular}{@{}lll@{}}
\toprule
Difference & 95\% CI & \textit{p}-value \\ \midrule
default Music- no Music & [-0.12, 0.024] & .19\\
custom Music - no Music & [-0.15, 0.0055] & .068\\
custom Music- default Music & [-0.10, 0.051] & .52\\
\bottomrule
\end{tabular}
\end{table}

Our third research question was whether music can evoke stronger emotions when watching data videos. We hypothesized that in this study, our data videos with music (default and custom) would lead to stronger reported emotions.

We computed the mean emotion intensity score for each group as well as pairwise differences between group means, with their 95\% CIs and $p$-values from independent samples $t$-tests. 
It can be seen from \autoref{fig:emotion}-(a) that the mean emotion intensity is moderate (close to 0.5 on a 0--1 scale) across all conditions. Looking at the pairwise comparisons reported in \autoref{fig:emotion}-(b) and \autoref{table_example_emotion}, contrary to what we expected, we have some evidence that no music led people to report more intense emotions than custom music. It is also possible that no music led them to report more intense emotions than default music, but the evidence is much weaker. Meanwhile, we did not find evidence of a difference between custom music and default music.

Conclusion: \textbf{The evidence goes against H3}. Not only do we lack evidence that music evokes more emotions, but we also have some evidence to the contrary.

\subsection{Additional Analyses}

The analyses in this subsection were not pre-registered.

\paragraph{\textbf{Message Emphasis.}}

\begin{figure}
    \centering
    \includegraphics[width=1\columnwidth]{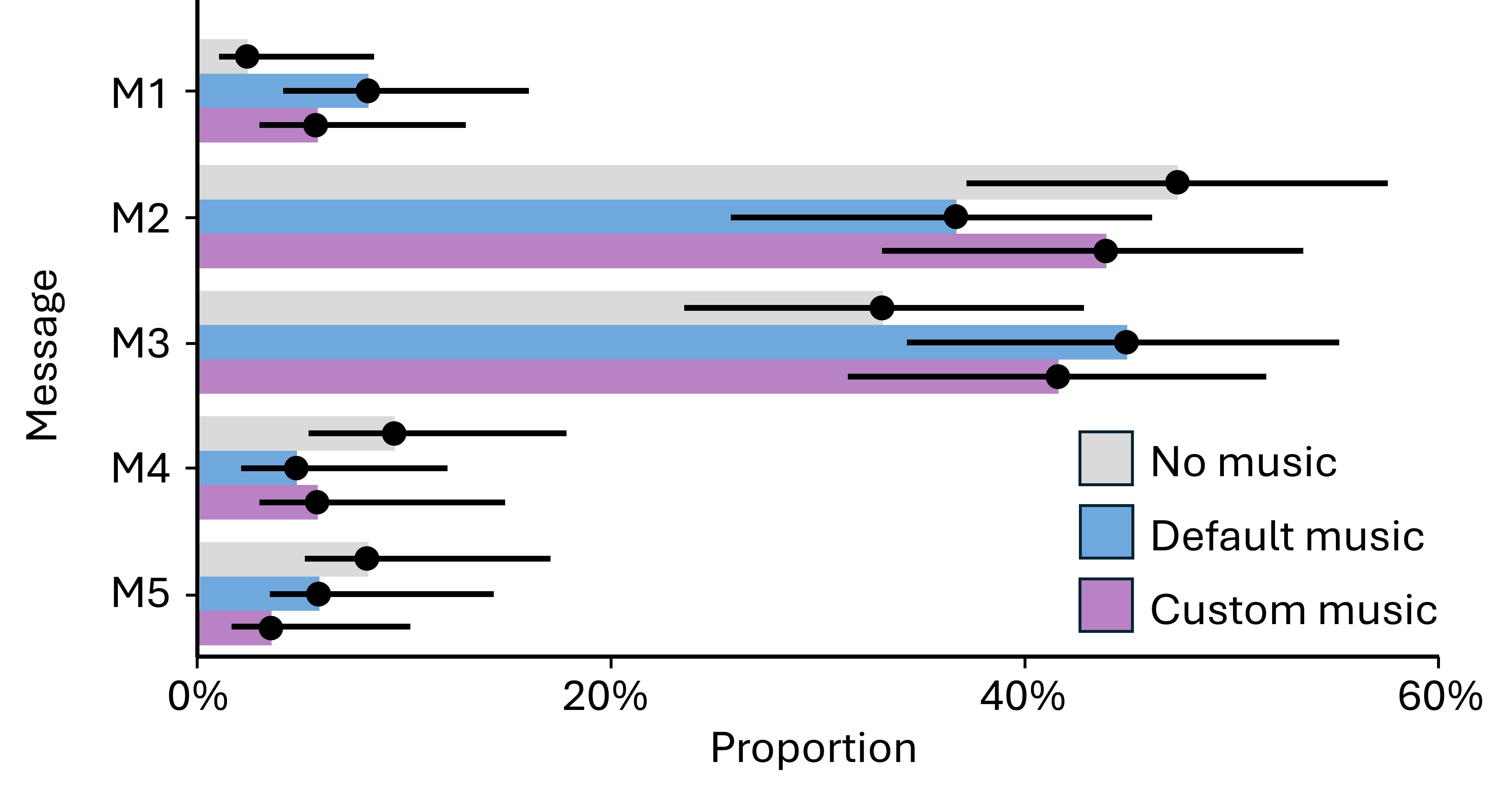}
    \caption{Proportion of participants who found that a particular message (among M1--M5) was the best communicated. Error bars are 95\% CIs.}
    \label{fig:music1}
\end{figure}

\begin{figure}
    \centering
    \includegraphics[width=1\linewidth]{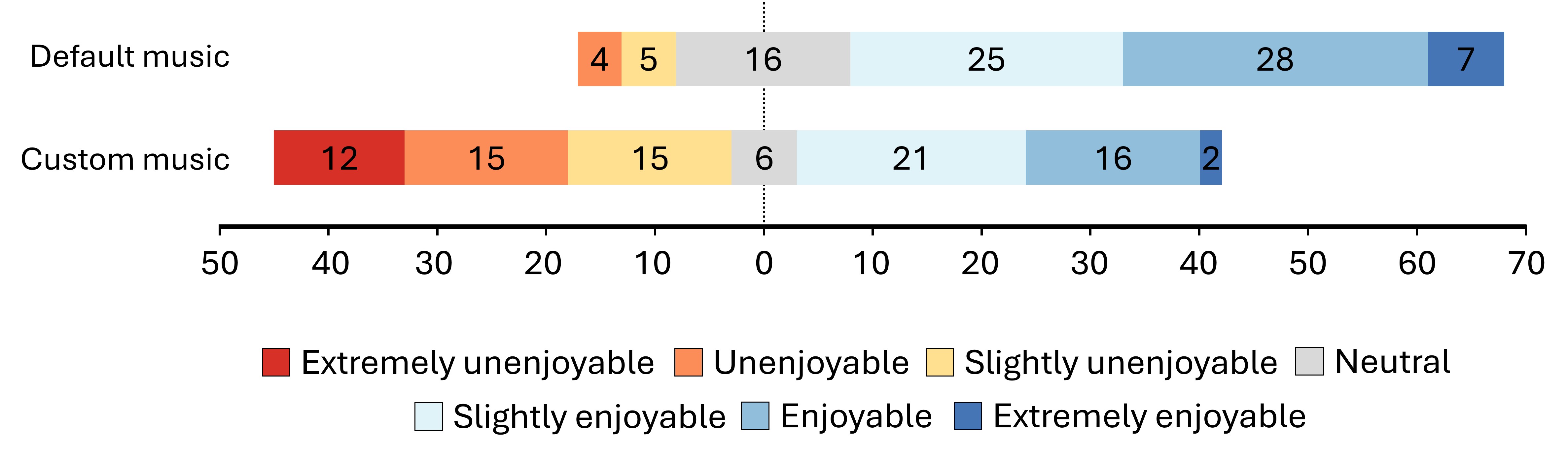}
    \caption{Comparison of how enjoyable the music was rated for the default music and custom music group.}
    \label{fig:Likert Scale}
\end{figure}

To answer our secondary research question about whether music can be used to emphasize particular messages (see \autoref{sec:message-emphasis}), we looked at what messages people thought were best communicated among the five we listed (see \autoref{tab:multinominal_messages} for the full list). In each condition, we computed the proportion of each response with simultaneous confidence intervals for multinomial proportions. Results are reported in \autoref{fig:music1}. 
Given the large overlaps in the CIs, we did not find evidence for a difference between the three conditions.

\paragraph{\textbf{Enjoyment of the Music.}}

\begin{figure}
    \centering
    \includegraphics[width=1\linewidth]{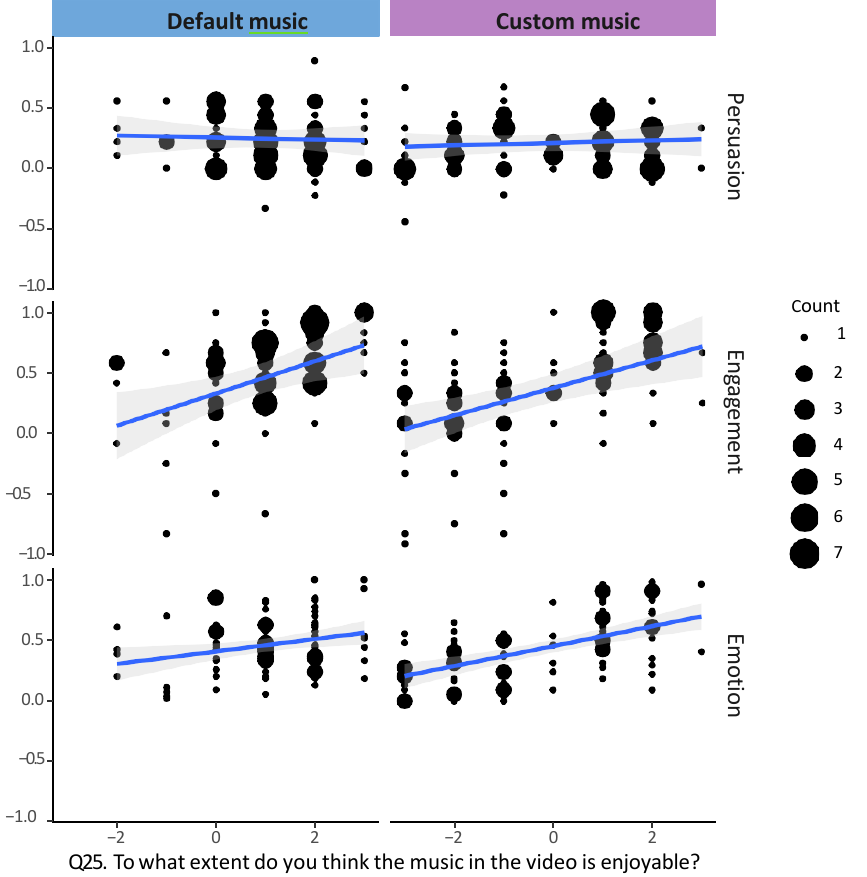}
    \caption{Correlation between participants' reported enjoyment of music (x-axis) and persuasion (top row), engagement (middle row), and emotion intensity (bottom row), both for default music (left column) and for custom music (right column).}
    \label{fig:correlation}
\end{figure}

\autoref{fig:Likert Scale} reports the extent to which participants reported enjoying the music in the default music and the custom music groups. 
It can be seen that participants' opinions about custom music were clearly more diverse than default music, with an overall preference for default music. We looked at whether there is a correlation between how much participants reported enjoying the music and their response to our three primary variables. Results are reported in \autoref{fig:correlation}. We found no evidence for a correlation between music enjoyment and attitudinal persuasion (top row in the \autoref{fig:correlation}). However, we found clear evidence for a correlation between music enjoyment and engagement (middle row in the \autoref{fig:correlation}), and between music enjoyment and emotion intensity (bottom row in the \autoref{fig:correlation}). The correlation seems particularly evident for engagement.
As we mentioned before, custom music was more polarizing, with participants reporting a positive experience, but others reporting a negative experience. It can be seen in the middle-right scatterplot that negative opinions generally led to low engagement. This could explain why custom music led to less engagement overall, compared to default music (as reported in \autoref{sec:engagement}). A similar reasoning can be applied to the emotion intensity metric.

\paragraph{\textbf{Video Interactions Behavior.}}
We logged participants' pause and rewind interactions while they were watching the data video. Results are summarized in \autoref{tab:interaction_behavior}. It appears that participants in the custom music group made more pauses, especially compared to the default music group, even though their average pause duration was shorter. Note, however, that we summed up interactions across all participants, and there were slightly more participants in the custom music group (87 vs. 85 in the other two). On average, there were 0.6 to 0.7 interactions (pauses or rewinds) per participant, for all three groups.



\begin{table}[!t]
\renewcommand{\arraystretch}{1}
\caption{Total number of video interactions, summed up across all participants in each group}
\label{tab:interaction_behavior}
\centering
\begin{tabular}{cccccc}
\hline
\multicolumn{1}{p{1.6cm}}{\centering \small Condition} & \multicolumn{1}{p{0.8cm}}{\centering \# \small Pauses} & \multicolumn{1}{p{1cm}}{\centering \small Avg. Pause Duration (s)} & \multicolumn{1}{p{0.8cm}}{\centering \small \# Rewinds} & \multicolumn{1}{p{1cm}}{\centering \small \% People Who Paused} & \multicolumn{1}{p{1cm}}{\centering \small \% People Who Rewinded} \\
\hline
\small No Music & 30 & 14 & 33 & 12\% & 16\% \\
\small Default Music & 16 & 15 & 34 & 5\% & 12\% \\
\small Custom Music & 40 & 11 & 24 & 10\% & 7\% \\
\hline
\end{tabular}
\end{table}

\paragraph{\textbf{Other Measures of Persuasion.}} 

\begin{figure}
    \centering
    \includegraphics[width=\columnwidth]{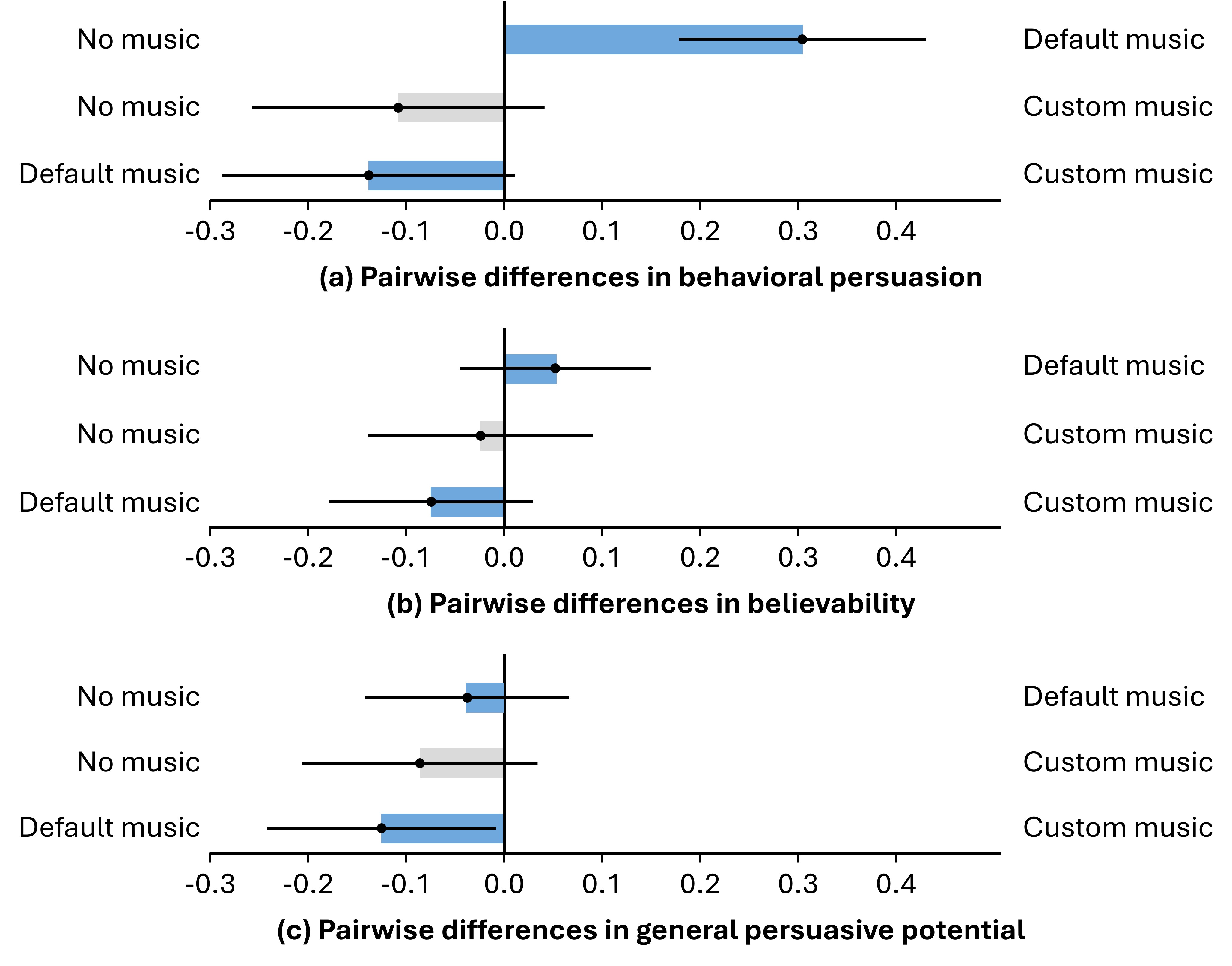}
    \caption{Pairwise differences in secondary measurements of persuasion (right condition minus left condition). Error bars are 95\% CIs.}
    \label{fig:secondary_pers}
\end{figure}

In addition to our primary measurement of persuasion, we measured behavioral persuasion, believability, and general persuasive potential as secondary measurements of persuasion (see details in \autoref{sec:measurements}). Results are reported in \autoref{fig:secondary_pers}. The results are overall consistent across the three metrics and with the attitude change metric (\autoref{fig:attitude_changes}). In particular, there is some evidence that default music outperformed custom music across all three metrics (bottom CI in the plots). There is also strong evidence that default music led to more behavioral persuasion than no music, but the evidence is inconclusive for the other two metrics. 

\begin{figure*}
    \centering
    \includegraphics[width=1\textwidth]{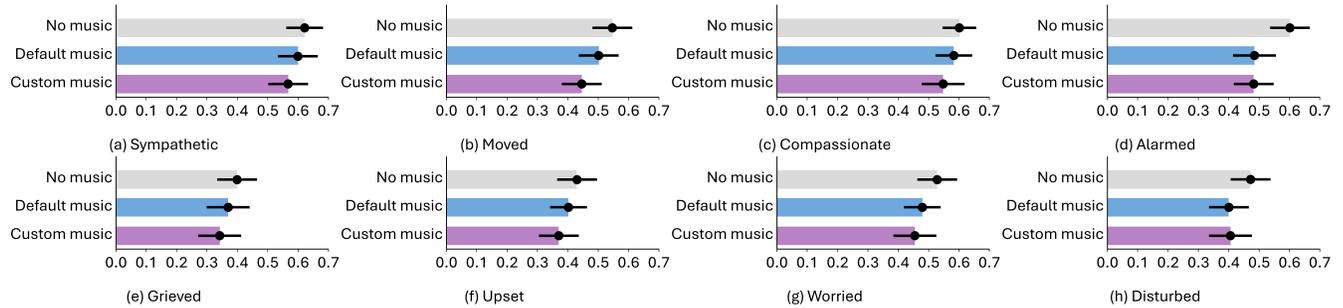}
    \caption{Reported mean intensity with 95\% confidence interval for all 8 emotions of Batson's list for three conditions. Emotion intensities are normalized between 0 and 1, with a score of 0 indicating no emotion, a score of 0.5 indicating moderate intensity, and a score of 1 indicating extreme intensity.}
    \label{fig:indi}
\end{figure*}

\paragraph{\textbf{Individual Emotions.}} 

Finally, we looked at the mean reported emotion intensity for each of the eight emotions individually (\autoref{fig:indi}). Reported intensities varied across emotions -- for example, participants generally reported medium intensities for the ``sympathetic'' and ``compassionate'' emotions, but lower intensities for the ``grieved'' emotion. In terms of differences between conditions, despite the overlaps in the confidence intervals, trends are consistent across all emotions and with our aggregated emotion intensity metric (\autoref{fig:emotion}): no music generally led to higher reported intensities, followed by default music and custom music. Therefore, it does not seem that our findings about emotion intensity are driven by a particular emotion, or that some emotions exhibit markedly different trends.

\section{Discussion}

We investigated a relatively uncharted territory, aiming to discern the influence of music on individuals' experiences with data videos across three dimensions: persuasion, engagement, and emotion. 
Our findings provide nuanced insights into the intricate relationship between music and the effectiveness of data-driven narratives, with results exhibiting variability across different dimensions. In this section, we delve into the implications of our findings, explore potential mechanisms underlying observed effects, and identify avenues for future work.

\subsection{The Persuasive Potential of Music in Data Videos}

In the context of data videos, prior research focused on visuals~\cite{garreton2023attitudinal} and affective narratives~\cite{sakamoto2022persuasive} to influence people's decision-making or takeaway of specific key messages. 
We add to this body of work by exploring how music affects the persuasive power of data videos. 
The findings from our study provide good evidence that data videos with music can lead to more persuasion, as measured through attitude change (\autoref{fig:attitude_changes}). A secondary measure of persuasion that captures the intention to act provides consistent results (\autoref{fig:secondary_pers}). 

However, we found stronger evidence for an effect with default music than with custom music, raising the possibility that the default music may have been more persuasive than our custom music. It is unclear from our findings which properties of the custom music may have reduced persuasion relative to the default music. One speculation is that while bespoke music has been found to be effective in traditional multimedia contexts such as prosocial activities~\cite{kniffin2017sound}, customers' purchasing decisions \cite{areni1993influence}, customer loyalty~\cite{morris1998effects}, and cognitive intents~\cite{martin2015effectiveness}, the effects of bespoke music in data videos are more difficult to predict.
It is plausible that more generic music (the default music) allows viewers to focus their attention on the facts and visuals more than the custom music, designed specifically to highlight the messages.

Our findings suggest new directions for future works to explore open questions on what type of music fits best in data videos and whether it is beneficial to customize music based on the data or story of the video. 
Interestingly, we found no evidence of a correlation between the level of enjoyment of the music and attitudinal persuasion, for both default music and custom music groups (\autoref{fig:correlation}-(a)). 
More research is needed to determine if there might be a mechanism at work similar to prior work that reported how music impacted people's implicit attitude toward consumer behavior in the commerce domain~\cite{johnson2011singing}. 
It will also be important to investigate the underlying mechanisms through which music influences viewer persuasion.

\subsection{Unanticipated Effects of Music on Engagement}

We expected that music would increase engagement (H2), yet we found no evidence that default music is more engaging than no music. 
Perhaps more surprisingly, we found evidence that custom music is \textit{less} engaging than no music (\autoref{fig:engagement}).
The effect may be driven by participants who did not like the custom music: more participants in the custom music group rated the music as not enjoyable than participants in the default music group (\autoref{fig:Likert Scale}), and those who rated the custom music as not enjoyable generally felt less engaged than those who enjoyed the music (\autoref{fig:correlation}-(b)).
These findings suggest that while the custom music we tested may have a less predictable effect on viewer engagement, the more generic default music offers a more consistent level of engagement and less diverse ratings of enjoyment of the default music.
A plausible conjecture is that the custom music may have been distracting to some participants. The custom music is likely more noticeable due to the music featuring more active elements to highlight key information and events, which is different from the more generic default music.
Such an interpretation is corroborated by prior studies where strong music was found to be disruptive, resulting in a lack of appeal~\cite{thompson2012fast}.
However, this is in contrast to prior work that suggests that music is generally beneficial for engagement~\cite{ferey2009multisensory}. 
Further research is needed to explore the underlying mechanisms driving these differences and to better understand how to design music suitable to positively influence viewers' engagement in data videos.

\subsection{Unanticipated Effects of Music on Emotion}


In our study, participants reported more intense emotions when no music was present in the data video (\autoref{fig:emotion}). 
This finding contrasts with prior work which suggested that music elicits emotional responses and influences viewers' affective  state~\cite{rogers2014music}.
Instead, we have evidence that our participants reported a lower intensity of emotions in the presence of music for both the default and custom music, and those who did not enjoy the music reported the lowest levels of emotions (\autoref{fig:correlation}--bottom).
It is worth mentioning that our measure of emotion intensity only captures participants' \textit{self-reported} emotions. It may be that music really made participants experience fewer emotions, but it could also be that music primed them to be more conservative in how they reported their emotions. The mechanisms behind this are however unclear. 
Additional studies are needed to investigate emotions while consuming data-driven content in the presence of music. 
These studies could include qualitative assessments beyond self-reporting, including follow-up questionnaires, interviews, or biomonitors in laboratory or real-world settings to identify the potential effect of music on participants' emotional states. 



\subsection{Music in Data Videos: Future Considerations}

Collectively, our analyses revealed varied outcomes regarding the influence of music on data videos across three dimensions -- persuasion, engagement, and emotion.  
While previous research underscores music's ability to captivate audiences and stir emotion, our findings do not consistently support these assertions, at least in the context of the data video and music we used in our study. 

Further studies could explore how to integrate music as a supplementary tool to measure comprehension of visual representations~\cite{krygier1994sound} or add audio cues (such as auditory icons or melodies) in narrative transitions to provide users with more content-dependent and intuitive information and reduce cognitive visual loads~\cite{tsuchiya2015data}. It remains unclear how far music can be used to emphasize messages. More collaboration between designers and musicians is needed to explore and better understand how music may be able to make data videos more effective. One challenge in composing music for data videos is the varied preferences and subjectivity in music. Data videos share that challenge with film music. Future studies could explore several alternatives and variations of music, including different melodies, genres, and styles of music to enhance (or purposefully disrupt) the narrative flow of the messages.  

While our results work as an initial assessment of the effect of music in data videos, future work may investigate the nuanced relationships and dichotomies between audio-visual elements in data videos. 

\section{Limitations and Future Work}
As with all studies, our work has limitations. 
First, while we followed a rigorous procedure to identify a suitable data video, our study used a single data video. Therefore, our results may not generalize to data videos across various topics and designs. More studies are needed to verify if our findings can be replicated across different data videos. Future studies could include multiple videos and use hierarchical models to disentangle the effects of video from the effects of music.

Second, we tested a single custom music track. While this music was created by a professional composer, other music designs may have been more effective. 
To account for varied preferences and subjectivity involved in music choices, future work should explore multiple alternatives and variations, including different melodies, genres, and styles of music. Additional avenues to explore include other types of sound such as data sonification~\cite{tsuchiya2015data} or algorithmically generated music~\cite{doornbusch2010algorithmic}. 


We only considered three measures -- persuasion, engagement, and emotion -- and music for data videos could have a range of other relevant effects on people 
including comprehension, distraction, implicit persuasion, actual behavior (beyond action \textit{intention}), and a broader range of emotions. 
Based on possible directions instigated by our research, future research could incorporate these measurements to provide a more comprehensive understanding of the impact of music on viewers' experiences of data videos.

Finally, our results may have been impacted by our participant pool. 
We recruited participants from Amazon Mechanical Turk, who, while fairly diverse, may not be representative of data video viewers. Our study was also conducted in a controlled setting, and participants were given limited choices regarding music selection. 
Individual preferences and contextual factors outside the experimental environment may influence viewers' responses to music in data videos. 

In the spirit of research transparency and of facilitating any replication and follow-up studies on this topic, we made all our study material available on OSF.

\section{Conclusion}

Our study sheds light on the underexplored realm of music's influence on the viewer's experience of data videos. To facilitate our investigation, we considered three conditions based on our selected data video on the global water crisis -- (1) no music, (2) default music, and (3) custom music to examine persuasion, engagement, and experienced emotions. We found that the default music made the video more persuasive, offering empirical support for its efficacy in enhancing viewers' attitudinal shift. 
However, we did not find any evidence that default music is more engaging than no music. 
Furthermore, our custom music yielded unexpected results, which underscores the nuanced role of music in shaping viewers' engagement and eliciting emotional responses. 
Contrary to expectations, the absence of music seems to have led people to report more intense emotions.  
Our study highlights the complex intersection of audio and visual stimuli in data videos and challenges the universal assumption that music enhances the viewer's experience. 
Further studies are needed to better understand the optimal use of music in data-driven 
 for engaging viewers or evoking emotions. Our study contributes insights into the intricate interplay among music, data visualization, and viewers' experience and offers avenues for future exploration and innovation in data-driven storytelling.


\balance



\IfFileExists{template.bbl}{%
  \input{template.bbl}
}{%
  \bibliographystyle{abbrv-doi-hyperref}
  \bibliography{template}
}

\end{document}